\documentclass[aps,preprint,preprintnumbers, amsmath, amssymb, prb,
showpacs,superscriptaddress]{revtex4-1}
\usepackage{times}
\usepackage{graphicx}
\usepackage{psfrag}
\usepackage{ae}
\usepackage{amsmath,amssymb}
\usepackage[usenames]{color}
\usepackage{float}

\begin{document}

\author{Jos\'e Rafael Bordin} 
\email{josebordin@unipampa.edu.br}
\affiliation{Campus Ca\c capava do Sul, Universidade Federal
do Pampa, Av. Pedro Anuncia\c c\~ao, 111, CEP 96570-000, 
Ca\c capava do Sul, RS, Brazil}

\author{Leandro B. Krott} 
\email{leandro.krott@ufrgs.br}
\affiliation{Campus Ararangu\'a, Universidade Federal de Santa Catarina, Rua 
Pedro Jo\~ao Pereira, 150, CEP 88900-000, Ararangu\'a, SC, Brazil}

\author{Marcia C. Barbosa} 
\email{marciabarbosa@ufrgs.br}
\affiliation{Instituto 
de F\'{\i}sica, Universidade Federal
do Rio Grande do Sul, Caixa Postal 15051, CEP 91501-970, 
Porto Alegre, RS, Brazil}

\title{Self-Assembly and Waterlike Anomalies in Janus Nanoparticles}

 \date{\today}

\begin{abstract}

We explore the pressure versus temperature phase diagram of dimeric Janus nanoparticles using Molecular Dynamics simulations.
The nanoparticle was modeled as a dumbbells particle, and have one monomer
that interacts by a standard Lennard Jones potential and another monomer that
is modeled using a two-length scale shoulder potential. Monomeric and dimeric systems
modeled by this shoulder potential show waterlike anomalies, and we investigate if a
Janus nanoparticle composed by one anomalous monomer will exhibit anomalous behavior
and self-assembly structures.
The influence of the non-anomalous monomer in the dimeric system properties 
was explored. We show that the diffusion anomaly is maintained, while the density anomaly
can disappear depending on the non-anomalous monomer characteristics . 
As well, the self-assembled structures are affected. Our results
are discussed in the basis of the distinct monomer-monomer interactions and on the two-length scale
fluid characteristics.

\end{abstract}

\maketitle
\section{Introduction}

The understanding the structural and thermodynamical behavior of 
colloidal systems is
important due to their  applications in medicine, catalysis, 
production
of photonic crystals, stable emulsions and other 
materials~\cite{Talapin10,ElL11, TuP13}.
The self-assembly structures of colloidal systems are very important 
to understand biological and chemical systems, like proteins and solutions~\cite{WaM08}. 
Experimental and computational researches show that Janus spheres present cluster 
formation and properties different from bulk systems~\cite{HoC08, ChW11}. When structured in
equilibrated aggregates, the charge asymmetry of each particle is preserved~\cite{HoC08}.

One of the relevant characteristics  of colloidal solutions is
the formation of stable self-assembly structures 
not present in traditional molecular systems~\cite{Boal00,Glotzer04}. 
The variety of the length scales and geometry
of the self-assemblies originate
from the different types of 
potential energies involved and from the shapes of colloidal particles such as
disk, spheric, rod-like and dumbbells.

The dumbbell shape is  quite particular. Each dumbbell is a dimer formed 
by two spheres with the same diameter 
that overlap with a separation that varies from an almost total overlap
to one or two monomer diameters. If each monomer of 
the dumbbell interacts with the other monomers by a one lenght scale
potential such as the 
Lennard Jones potential the pressure
temperature phase diagram resembles the diagram of the monomeric case~\cite{Oliveira10}.
If, however, each monomer of one dumbbell interacts
with the other dumbbell by a two length scale potential, 
the symmetry breaks and a phase in which the  dimers are aligned similar
to a liquid-crystal phase appears~\cite{Oliveira10}. The dumbbells diffuse
along this line while no transport between the lines is observed.
 One special type
of dummbbell is the  Janus particles~\cite{Yin01}. They 
are
characterized by having two types of monomers. For instance
one hydrophobic and the other 
hydrophilic. In this case the breaking
of the spherical symmetry leads to a more complex set of phases. The hydrophilic monomers 
attract each other while the hydrophobic monomers repeal forming 
lamellar and micellar agregates of different length scales. The presence
of this type of structures is not new in the literature.
Systems with competing interactions, first neighbors repulsion and second neighbors
attraction exhibit these same lamellar and micellar formations~\cite{Bagdassarian88,Barbosa90} as observed in the Janus particles. One particular example where this type of interactions appear is in
the mixture of water, oil and surfactant~\cite{Barbosa90}.

Recently, nanocomposites of dumbbells and Janus-type particles 
have been synthesized in large scale using several 
techniques, as physical vapor 
deposition~\cite{SiC14}. These asymmetric colloidal dimers can be obtained through 
deposition of gold nanoparticles~\cite{Lu02, YoL12}, 
phase separation and immobilization in liquid-liquid or liquid-gas 
interfaces~\cite{HuZ12}, using the  interfacial activity of reverse micelles 
and microemulsions~\cite{Li99}
and other techniques.

Due to the resemblance between Janus particles and competing
interaction systems, Janus dumbbells behave
 as surfactant in water-based emulsions due its amphiphilic 
properties~\cite{SoK11, TaI05, Liu13}. It has been shown that Janus particles are efficient 
in hydrophobization of textile materials, that depends on the size of the 
particles~\cite{SyK11}. By a simple geometrical change, varying from spherical 
symmetric to cylindrical structural shapes, particles with new properties can be created~\cite{WaM08}.

Simple models found rich pressure versus temperature phase diagrams and critical behavior of dumbbells 
and charged colloids~\cite{MiD13,BiL14}. Varying the interaction potential, lamellar 
structures and gas-liquid phase separation are found in Janus particles~\cite{MuT14,MuC13} in a very similar facion as in the competing
interaction models these phases were found in the past~\cite{Barbosa90}.
Micellization and phase separation also are related in such models~\cite{ChR11}. Homogeneous 
crystal nucleation of colloidal hard dumbbells can be suppressed by high free energy barrier 
or slow dynamics~\cite{NiD11}. Controlling the attractive of patchy colloids, liquid-gas 
separation can be suppressed and a gel phase can be formed~\cite{BiB11}.

Recently, the production of silver-silicon (Ag-Si) hybrid Janus dimmers was reported~\cite{SiC14}.
While silver is a material that do not shows anomalous behavior, silicon is classified as a anomalous fluid.
Anomalous fluids exhibit a set of properties called anomalies that divert from the observed
in simple fluids. The increase of density with the temperature at a fixed pressure
and the increase of diffusivity under compression are examples of 
these anomalies. Water is the most well known fluid that
present thermodynamic, dynamic and structural anomalous 
behavior~\cite{Ke75,An76,Pr87}, with 72 known anomalies~\cite{URL}. 
In addition, silicon~\cite{Sa03} and others material, as silica~\cite{An00,Sh02,Sh06},
Te~\cite{Th76}, Bi~\cite{Handbook}, 
Si~\cite{Sa67,Ke83}, $Ge_{15}Te_{85}$~\cite{Ts91},  liquid 
metals~\cite{Cu81}, graphite~\cite{To97} and 
$BeF_2$~\cite{An00} shows thermodynamic anomalies~\cite{Pr87}, while
silicon~\cite{Mo05} and silica~\cite{Sa03,Sh02,Sh06,Ch06} 
show a maximum in the diffusion coefficient at constant temperature.
These systems and their density and diffusion anomalies can
be described in an effective way by a two
length scale core-softened potential. This type of 
coarse graining model when used in dumbbell particles 
in which both monomers are identical and interact by two
length scale potential
shows the presence of liquid-crystal type of order discribed above~\cite{Oliveira10}. 

Naturally  given 
 the production of silver-silicon janus the question arises what happens when a core-softened 
potential particle  and an hydrophobic or an hydrophilic particles particle
are jointed to form a dumbbell. Is the micellar Janus
behavior dominant or is the liquid-crystal observed in two length
scales dummbbells the main behavior? In order to answer
to this question in this paper we study the two limits of 
this question. The first system is a dumbbell in which one monomer interact
through a core softened potential with a very small attractive part and 
another monomer that is purely repulsive. The second system
is a dumbbell in which one monomer interacts by a two length scales
potential but the second shows attraction.

The paper is organized as follows: first we introduce the model and describe the methods 
and simulation details; next the results and discussion are given; and 
then we present our conclusions.

\section{The Model and the Simulation details}
\label{Model}

 The system consists of $N$ dimmeric nanoparticles, in a total of $2N$ 
monomers. 
 The Janus dumbbells was modeled using two spherical 
symmetric particles, each one with mass $m$ and both with an
 effective diameter $\sigma$, linked rigidly at a distance 
$\lambda = 0.8$. Three types of monomers were used.
 Monomers of type A interact with another
monomer of type A through a core-softened (CS)
pontential, while monomers of type B interact with another
monomer B with a 
 purely repulsive potential and 
 monomers of type C interact with
another monomer C  through an  attractive Lennard-Jones (LJ) potential. The two length 
scale potential that gives to the monomeric 
and dimeric 
system made only with  particles A the
 anomalous characteristics is  defined 
as~\cite{Oliveira06a}
 \begin{equation}
 \frac{U^{AA}(r_{ij})}{\varepsilon} = 4\left[ \left(\frac{\sigma}{r_{ij}}\right)^{12} -
 \left(\frac{\sigma}{r_{ij}}\right)^6 \right] + u_0 {\rm{exp}}\left[-\frac{1}{c_0^2}\left(\frac{r_{ij}-r_0}{\sigma}\right)^2\right]
 \label{AlanEq}
 \end{equation}
 where $r_{ij} = |\vec r_i - \vec r_j|$ is the distance between two A 
particles $i$ and $j$.
 This equation has two terms: the first one is the standard 12-6 LJ
 potential~\cite{AllenTild} and the second one is a Gaussian
 centered at $r_0$, with depth $u_0$ and width $c_0$.
 Using the parameters $u_0 = 5.0$, $c = 1.0$ and $r_0/\sigma = 0.7$ 
this equation 
 represents a two length scale potential, with one scale 
 at  $r_{ij}\equiv r_1\approx 1.2 \sigma$, where the 
 force has a local minimum, and the other scale at  
$r_{ij}\equiv r_2 \approx 2 \sigma$, where
 the fraction of imaginary modes has a local minimum~\cite{Oliveira10}. The 
cutoff radius
 for this potential is $r_c = 4.0$. Despite the mathematical simplicity, 
 monomeric and dimeric fluids using this shoulder model 
   does exhibit the density, the diffusion, and the
response functions anomalies observed in water and other 
anomalous fluids~\cite{Oliveira06a, Oliveira06b, Oliveira10b, Kell67,Angell76}.

 \begin{figure}[ht]
 \begin{center}
 \includegraphics[width=8cm]{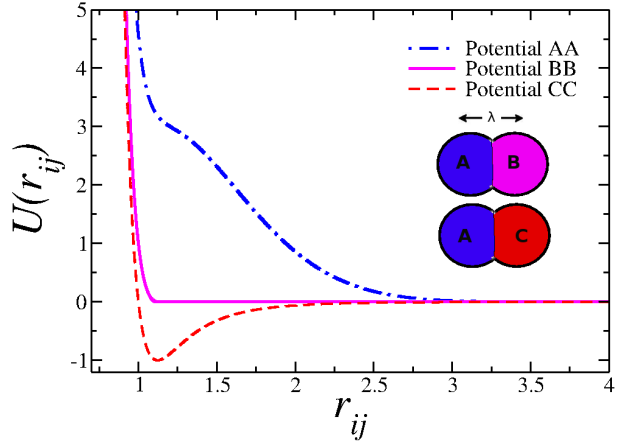}
 \end{center}
 \caption{Interaction potentials used in our simulations: the 
shoulder potential AA (dot-dashed blue line), the 
 CSLJ potential BB with $r_c = 2^{1/6}$ (solid magenta line) and the 
 CSLJ potential CC with $r_c = 2.5$ (dashed red line).
 Inset: Janus nanoparticles formed by A-B monomers and by A-C monomers.}
 \label{fig1}
 \end{figure}
 
 The repulsive interactions are given by a cut and shifted 
Lennard-Jones (CSLJ)  potential while the attractive 
particles interact by Lennard-Jones potential namely
 \begin{equation}
 \label{LJCS}
 U^{\rm{CSLJ}}(r_{ij}) = \left\{ \begin{array}{ll}
 U_{{\rm {LJ}}}(r_{ij}) - U_{{\rm{LJ}}}(r_c)\;, \qquad r_{ij} \le r_c\;, \\
 0\;, \qquad \qquad \qquad \qquad \quad r_{ij}  > r_c\;.
 \end{array} \right.
 \end{equation}
 \noindent Here, $U_{{\rm {LJ}}}$ is the standard 12-6 LJ potential, included 
in the first term of equation~(\ref{AlanEq}),
 $r_c = 2^{1/6}$ is the cutoff for the B-B interaction, namely 
potential BB, and $r_c = 2.5$ for the C-C interaction, 
 namely potential CC. The first case model a purely repulsive 
excluded volume interaction, 
 while the second one has a attractive well. The interactions
 between A-B, B-C and A-C monomers are also
purely repulsive   given by the equation~(\ref{LJCS}) 
with $r_c = 2^{1/6}$.
 The potentials are shown in figure~\ref{fig1}. The internal bonds between 
each dimer remain fixed using the 
 SHAKE algorithm~\cite{Ryc77}.

 Here we explore two model systems. In the first model the nanoparticles are 
composed by one monomer of type A and one monomer
 of type B, representing the combination 
of an anomalous fluid with an hydrophobic system.  In the second case, the dumbbell is modeled by a monomer 
of type A and one monomer of type C representing a janus
particle in which one monomer is an anomalous fluid and 
the other is a very hydrophilic system.
Both cases are illustrated in the inset of the figure~\ref{fig1}. 
 
Molecular dynamics simulations is used in order to obtain the pressure 
versus temperature ($p\times T$) phase diagram.
The simulations were performed in the canonical ensemble using the 
ESPResSo package~\cite{espresso1, espresso2}.
A total number of 1000 particles (500 dimers) where used. The number density
is defined as $\rho = N/V$, where $V=L^3$ is the volume of the cubic 
simulation box.
Standard periodic boundary conditions are applied in all directions. 
The system temperature was fixed using the Langevin thermostat 
with $\gamma = 1.0$,
and the equations of motion for the fluid particles were integrated
using the velocity Verlet algorithm, with a time step $\delta t = 0.01$. 
We performed $5\times10^5$ steps to equilibrate the system. 
These steps are then followed by $5\times10^6$ steps for the results 
production stage. 
To ensure that the system was equilibrated, the pressure, kinetic energy
and potential energy were analyzed as function of time, as well several 
snapshots at distinct simulation times.
To confirm our results, in some points we carried out simulations with 
2000 and 5000 particles, and 
essentially the same results were observed.  
    
To study the dynamic anomaly the relation between 
the mean square displacement (MSD) with time is analyzed,  namely
\begin{equation}
\label{r2}
\langle [\vec r_{\rm cm}(t) - \vec r_{\rm cm}(t_0)]^2 \rangle =\langle 
\Delta \vec r_{\rm cm}(t)^2 \rangle\;,
\end{equation}
where $\vec r_{\rm cm}(t_0) = (x_{\rm cm}(t_0)^2 + y_{\rm cm}(t_0)^2 
+ z_{\rm cm}(t_0)^2)^{1/2} $ 
and  $\vec r_{\rm cm}(t) = (x_{\rm cm}(t)^2 + y_{\rm cm}(t)^2 
+ z_{\rm cm}(t)^2)^{1/2} $
denote the coordinate of the nanoparticle center of mass (cm)
at a time $t_0$ and at a later time $t$, respectively. The MSD is
 related to the 
diffusion coefficient $D$ by
\begin{equation}
 D = \lim_{t \rightarrow \infty} \frac{\langle \Delta \vec r_{\rm cm}(t)^2 \rangle}{6t}\;.
\end{equation}

The structure of the fluid was analyzed using the radial distribution 
function (RDF) $g(r_{ij})$, and the pressure was evaluated with the 
virial expansion. 
In order to check if the Janus system shows density anomaly we evaluate the 
temperature of maximum density (TMD). Using thermodynamical relations, the
TMD can be characterized by the minimum of the pressure versus
temperature along isochores,
 \begin{equation}
  \left(\frac{\partial p}{\partial T}\right)_{\rho} = 0\;.
  \label{TMD}
 \end{equation}
\noindent The fluid and micellar region in the $p\times T$ phase diagram 
were defined analyzing the structure with $g(r_{ij})$,
snapshots and the diffusion coefficient $D$.

In this paper all the physical quantities are computed
in the standard LJ units~\cite{AllenTild},
\begin{equation}
\label{red1}
r^*\equiv \frac{r}{\sigma}\;,\quad \rho^{*}\equiv \rho \sigma^{3}\;, \quad 
\mbox{and}\quad t^* \equiv t\left(\frac{\epsilon}{m\sigma^2}\right)^{1/2}\;,
\end{equation}
for distance, density of particles and time , respectively, and
\begin{equation}
\label{rad2}
p^*\equiv \frac{p \sigma^{3}}{\epsilon} \quad \mbox{and}\quad 
T^{*}\equiv \frac{k_{B}T}{\epsilon}
\end{equation}
for the pressure and temperature, respectively, where $\sigma$ is the 
distance
parameter, $\epsilon$ the energy parameter and $m$ the mass parameter.
Since all physical quantities are defined in reduced LJ units, 
the $^*$ is  omitted, in order to simplify the discussion.

\section{Results and Discussion}
\label{Results}
   
\subsection*{A-B type nanoparticles}

 \begin{figure}[ht]
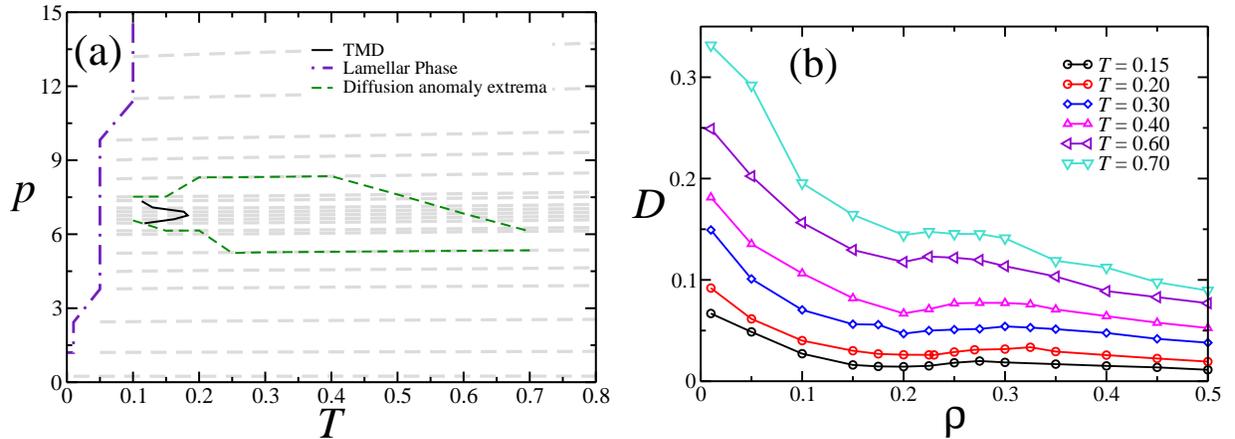

 \begin{center}
 \includegraphics[width=8cm]{fig2a.eps}
 \includegraphics[width=8cm]{fig2b.eps}
 \end{center}
 \caption{(a) $p\times T$ phase diagram for the system composed by A-B 
 Janus nanoparticles. The black line denotes the TMD line, the dashed green 
 line the diffusion anomaly region extrema and the dot-dashed 
 purple line the separation between the fluid and the lamellar phase. Grey 
 lines are the isochores. (b) Center of mass diffusion 
 coefficient $D$ as function of density $\rho$ for different temperatures, 
 showing the diffusion anomaly.}
 \label{fig2}
 \end{figure}
 
 We analyze a system of janus particles in
which one monomer is of type A and 
the other is of type B, the AB system. Previous 
works~\cite{Oliveira10b} 
 has shown that dimers with two monomers of type A exhibit thermodynamic,
 dynamical and structural waterlike anomalies and a 
liquid-crystal lamellar phase. 
 First, we focus in the thermodynamic (density)
 and dynamical (diffusion) anomalies.
 
 The $p\times T$ phase diagram of this
AB system is shown in the figure~\ref{fig2}(a). 
This system exhibits a region of pressures and temperatures
in which the density increases with the increasing density.
The temperature of maximum density for different pressures
is illustrated as a solid line in the figure~\ref{fig2}(a).
 The points where obtained using equation~\ref{TMD}.
 The diffusion coefficient versus density
for different temperatures is shown in the figure~\ref{fig2}(b). 
 For some densities and temperatures, the diffusion coefficient, $D$, increases with 
$\rho$ what characterizes an anomalous behavior.
 As a result for certain
temperatures there is a minimum and maximum diffusion coefficient. These
two points are illustrated as a bottom and an upper dashed lines
in the pressure versus temperature phase diagram in the figure~\ref{fig2}(a).  Therefore the
   the region of the 
diffusion anomalous behavior 
in the pressure versus temperature 
phase diagram englobes the region
of the density anomaly what is identified as
hierarchy of anomalies. This hierarchy
is also  observed for the monomeric~\cite{Oliveira06a} and A-A dimers~\cite{Oliveira10b}.

 \begin{figure}[ht]
 \begin{center}
 \includegraphics[width=4cm]{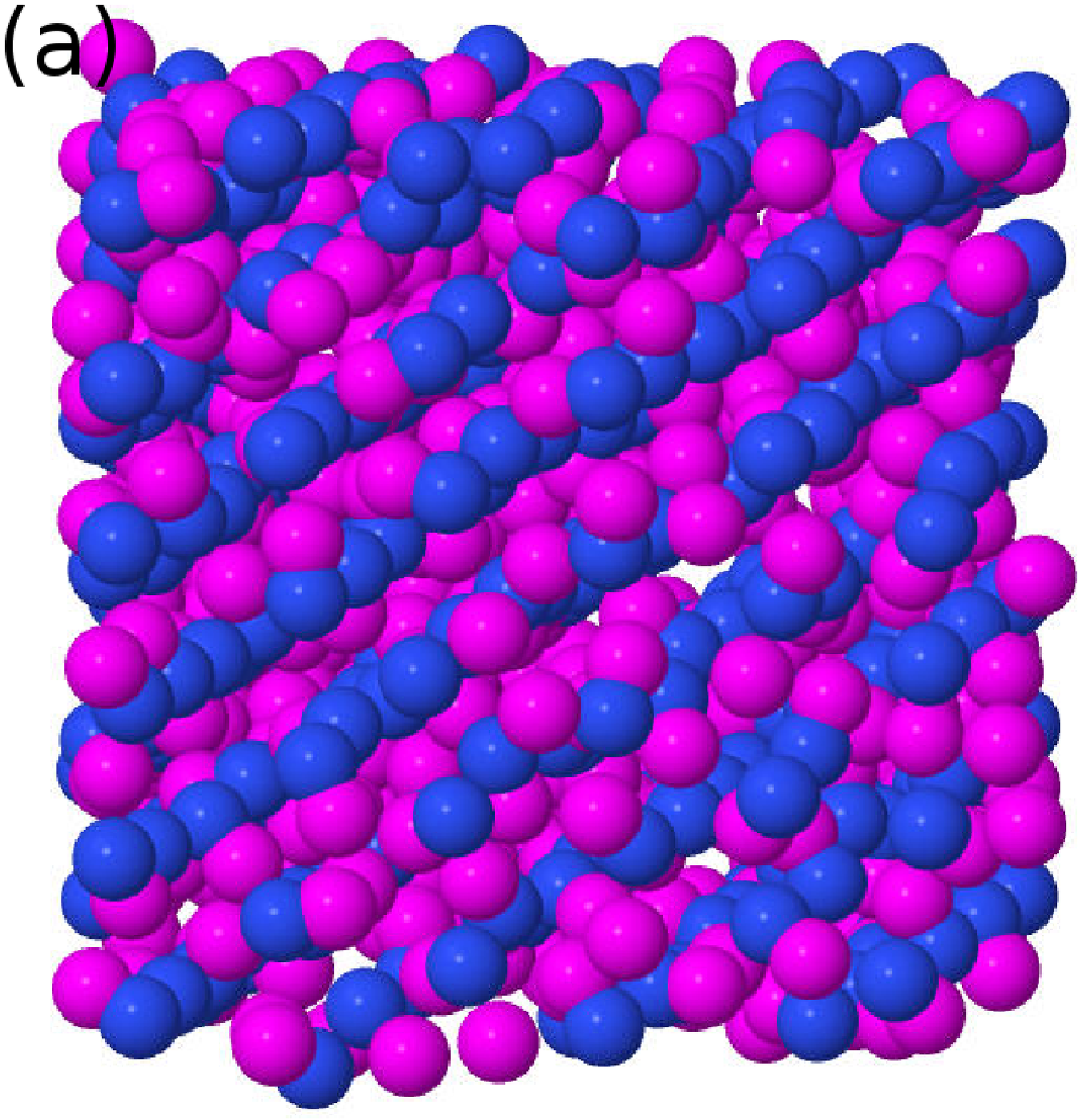}
 \includegraphics[width=4cm]{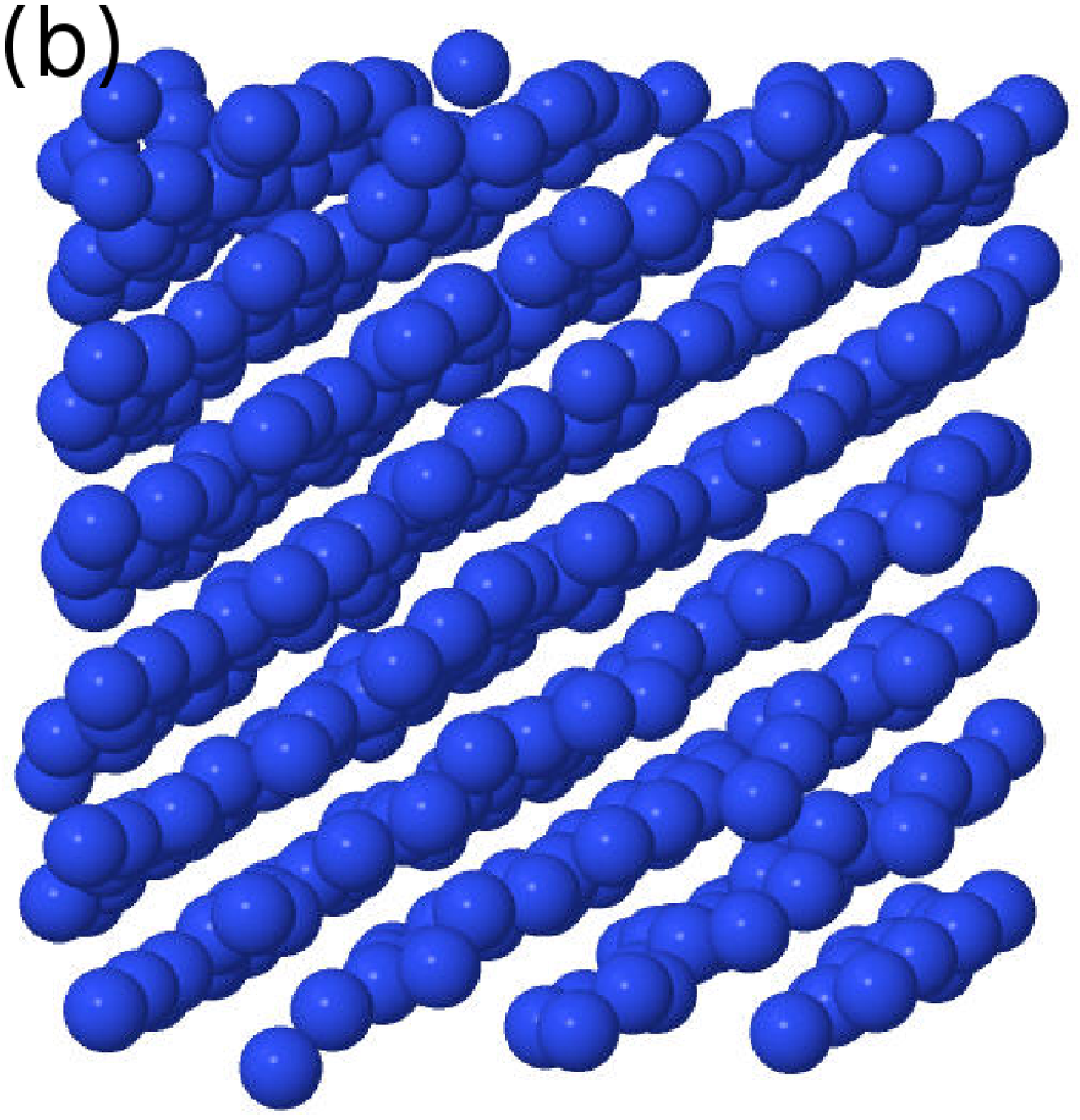}
 \includegraphics[width=4cm]{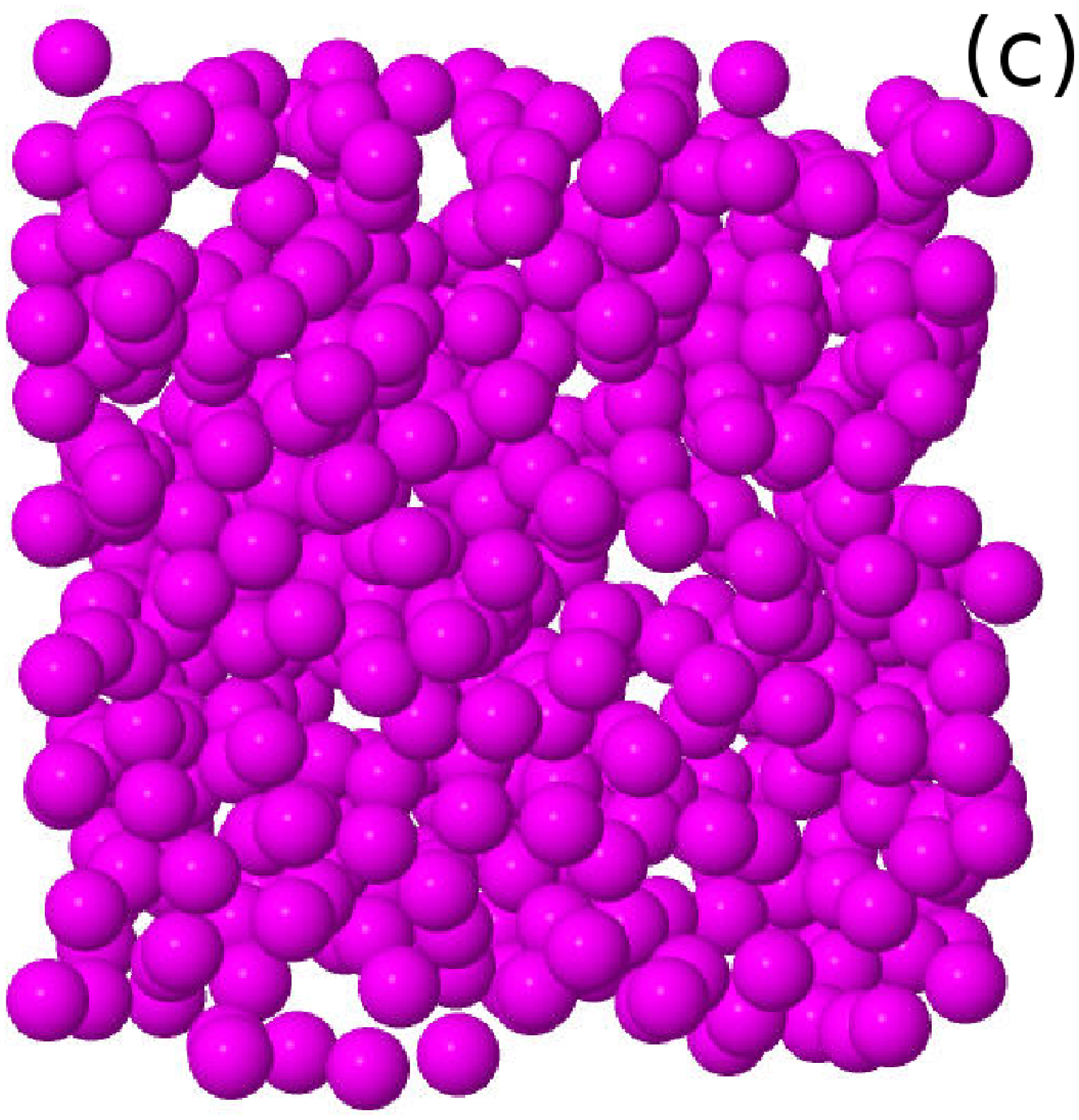}
 \includegraphics[width=6cm]{fig3d.eps}
 \end{center}
 \caption{(a)System snapshot in the lamellar phase, at $T = 0.1$ and $\rho = 0.4$ 
 including all particles, (b) only the A (blue) monomers and (c) only the B 
 (pink) monomers. (d) Radial distribution function for A-A pairs (solid black line)
 and B-B pairs (solid red line).}
 \label{fig3}
 \end{figure}

 The fluid-solid separation occurs at 
lower temperatures compared with the 
 A-A case~\cite{Oliveira10b}, mainly due the excluded volume by B monomers. 
 In the fluid phase, no ordering was observed. In the solid region, defined
 when the fluid is structured and with $D\approx0$, the system is structured
 in a lamellar micelle. The figure~\ref{fig3}(a) shows a snapshot of the 
 nanoparticles in this region, at $T = 0.1$ and $\rho = 0.4$. To clarify the structure
 we analyze the A and B monomers separately. Figure~\ref{fig3}(b) and (c) are snapshots
 of A monomers and B monomers, respectively. As we can see, the A monomers
 are in a lamellar  well defined structure, while the B monomers are disordered, 
 with a fluid-like behavior. Since the center of mass diffusion coefficient 
 is approximately null, this indicates that the A monomers are fixed in 
 the lammelar structure, while the  B monomers are spinning around the
 A monomers, in a fluid-like behavior. 
The radial distribution function 
 for this point, shown in figure~\ref{fig3}(d), reinforces this conclusion. 
 The RDF for A monomers is characteristic of a solid, and for
 B monomers is clear the RDF of a gas. For all densities and 
 temperatures simulated only this lamellar structure was observed 
 in the solid phase. The snapshots and RDF for this points are omitted
 for simplicity. 
 
 The  hard sphere-two length scales dumbbell forms
a plane of two length scales monomers as illustrated
in the figure~\ref{fig3}(b). In the case of
a dumbbell in which both monomers interact through a two
length scales potential the two monomers minimize the
free energy by being apart a distance $d\approx2\sigma$ between
the lines. Inside each line the distance is $d_1\approx1\sigma$

The reason for this difference is that
while the two length scales particles
minimize the free energy being either
at $d\approx2\sigma$ e $d_1\approx1\sigma$, the
hard core monomers are limited by
the hardcore distance. The combination
of these two particles that are linked
by the dumbbell leads to the
appearance of the planes.

Another effect from the Janus characteristic of the 
nanoparticle is that the liquid-crystal phase observed for A-A 
dimers was not observed in the A-B case.
Once the anomalies are preserved,
our results indicates that in the A-B case the two length scales potential
determines three fluid behavior.

\subsection*{A-C type nanoparticles}

 \begin{figure}[ht]
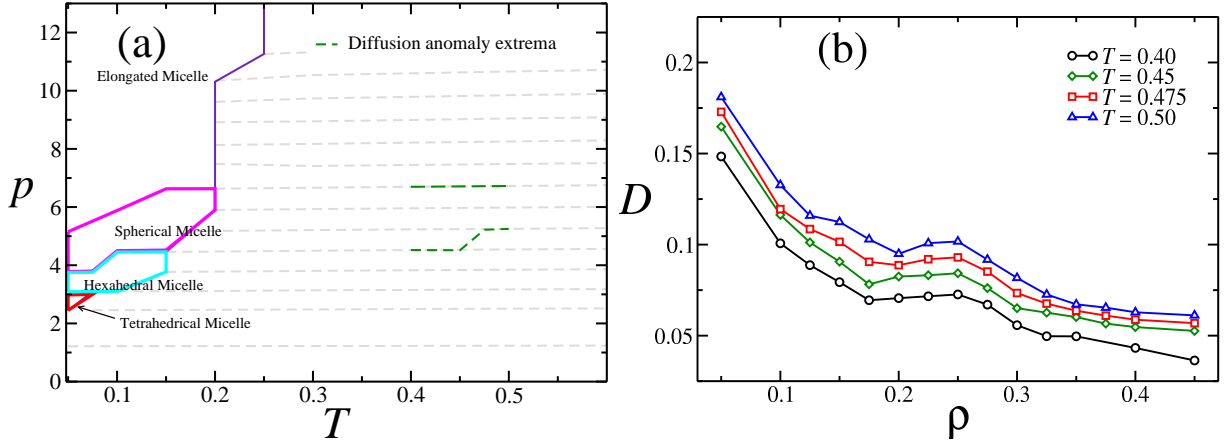

 \begin{center}
 \includegraphics[width=8cm]{fig4a.eps}
 \includegraphics[width=8cm]{fig4b.eps}
 \end{center}
 \caption{(a) $p\times T$ phase diagram for the system composed by A-C Janus nanoparticles. 
 The dashed green line the diffusion anomaly region extrema and the distinct micellar 
 regions are depicted in the graphic. Grey lines are the isochores. (b) Center of mass 
 diffusion coefficient $D$ as function of density $\rho$ for different temperatures, 
 showing the diffusion anomaly.}
 \label{fig4}
 \end{figure}

 Next, we analyze the A-C Janus dimers. Replacing the purely repulsive B monomer by 
 the attractive C monomer leads to changes in the pressure
versus temperature phase diagram when compared with the A-B case. 
As we show in figure~\ref{fig4}(a), the density anomaly disappears and the diffusion anomaly regions
shrinks. The diffusion anomalous increase as the system is compressed is shown in figure~\ref{fig4}(b) 
for isotherms between $T=0.4$ and $0.5$ -- the region where the anomaly was observed.
The waterlike anomalies can be related with the particles separation, 
at a higher distance with small energy, or at a smaller distance and higher energy - the two length
scales in the equation~(\ref{AlanEq})~\cite{Oliveira10}. The particles moves to the closest 
configuration, the first length scale, as we increase the temperature - or increase the 
entropic contribution to the free energy, while the enthalpic contribution to the free energy
is higher when the characteristic distance is the second length scale.
The competition between the two length scales, or between entropy and enthalpy, leads to the 
waterlike anomalies~\cite{Oliveira09}. Therefore, the suppression of the density anomalous region 
and the the occurrence of diffusion anomaly only at high temperatures are consequence 
of the attractive interaction, witch favors the enthalpic contribution to the free energy.

 \begin{figure}[ht]
 \begin{center}
 \includegraphics[width=10cm]{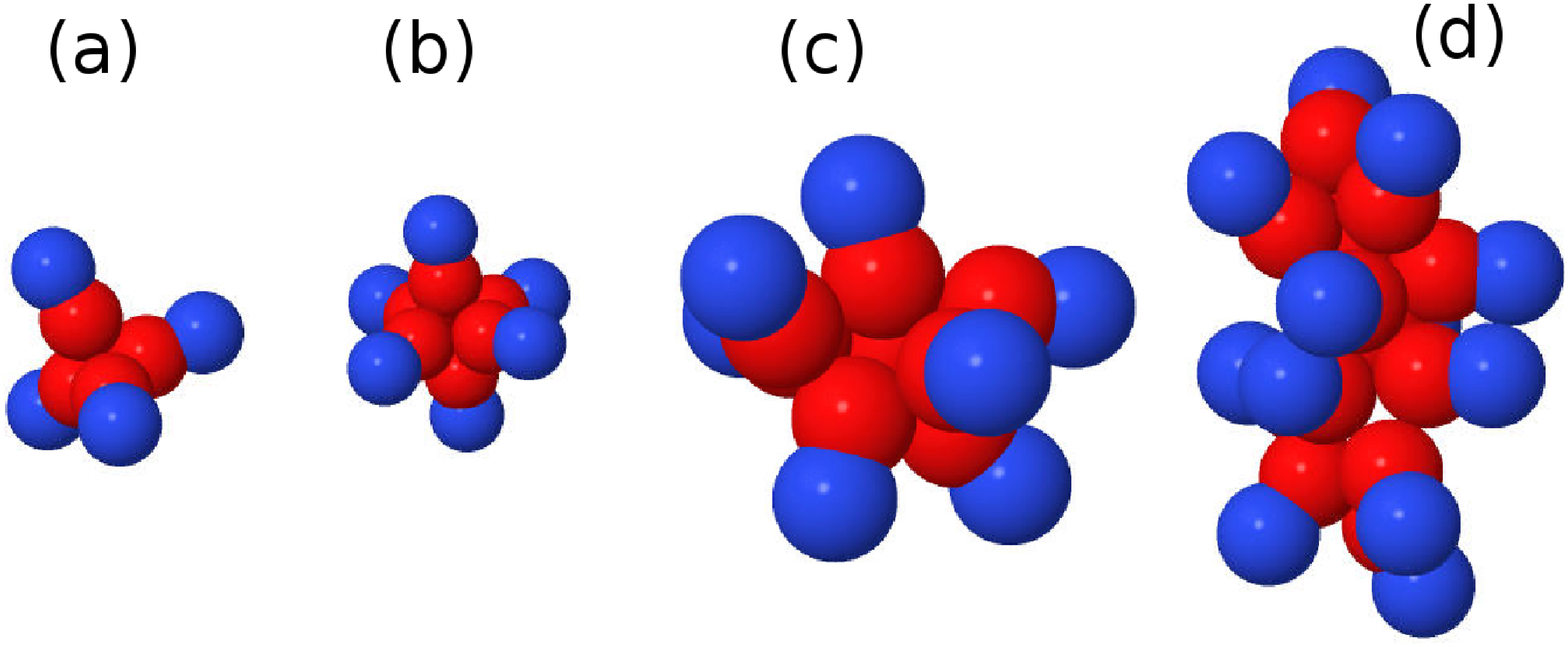}
 \includegraphics[width=6cm]{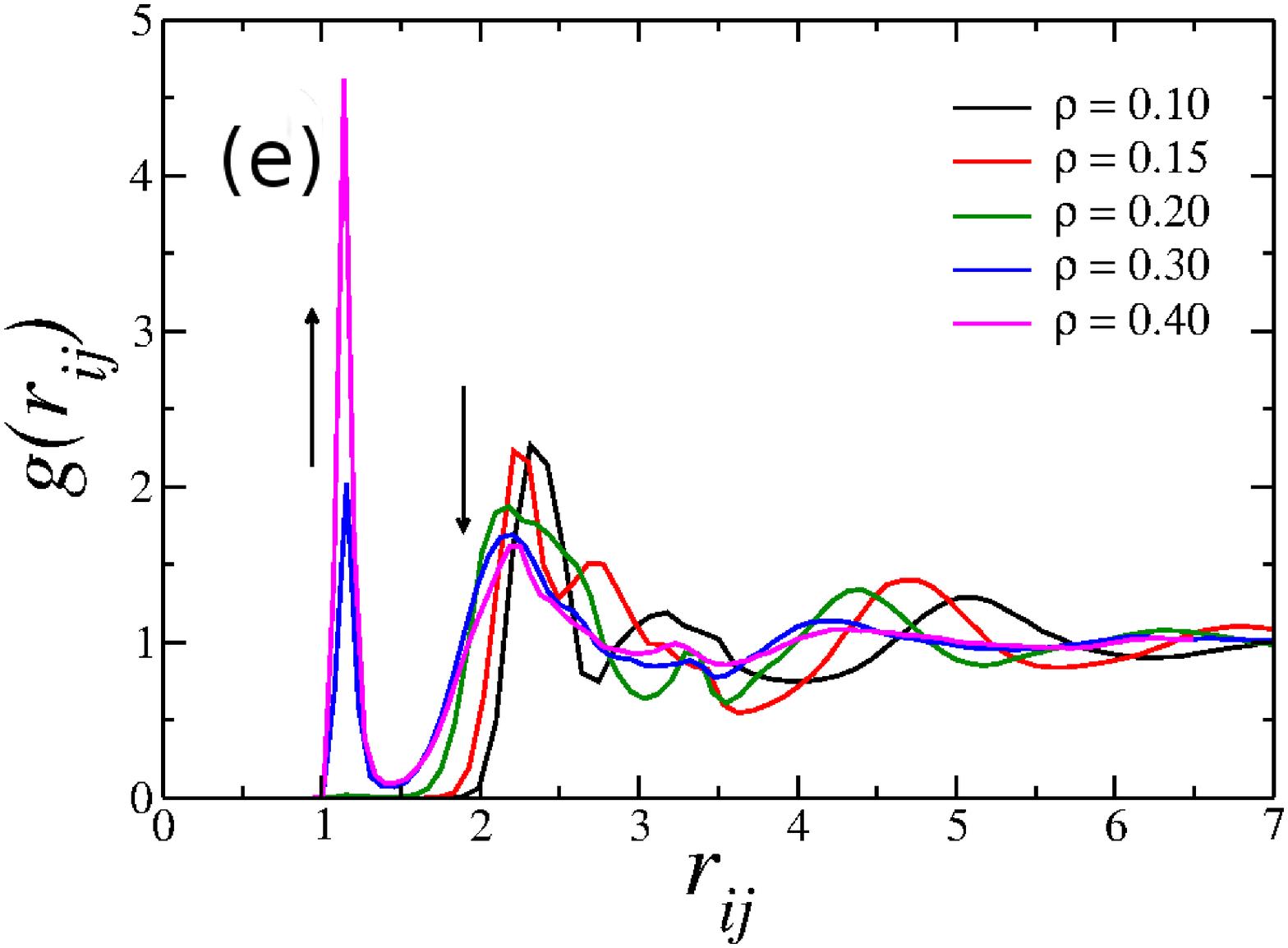}
  \end{center}
 \caption{System snapshot in the micellar phase, at $T = 0.05$ and (a) $\rho = 0.10$, (b) 0.15, 
 (c) 0.20 and (d) 0.30, respectively, showing the distinct self-assembled micelles: tetrahedral, 
 hexahedron, spherical and elongated, form left to right. (e) Radial distribution function 
 for A-A pairs at temperature $T = 0.05$ and for different densities, showing the competition
 between the length scales. The arrows indicates the behavior of each peak as the density increases.}
 \label{fig5}
 \end{figure}

 Another difference from the A-B nanoparticles are the self-assembled structures. 
 While for the first case only a lamellar phase was observed, 
 for A-C monomers we obtained four different micelles structures. 
 The micellation temperature is higher than in the previous case, as expected
 for attractive particles. To analyze the micellar phase we will discuss the
 temperature $T=0.05$, where all structures where
 observed. At small pressures, in the region limited 
 by the red line in the figure~\ref{fig4}(a),
 the dumbbells are structured in pyramidal 
 tetrahedral clusters, similar to the water tetrahedral cluster, 
 with the C monomers attached and pointing to the center
 of the cluster. For higher pressures, inside the region
 limited by the cyan line in the figure~\ref{fig4}(a), the nanoparticles 
 are in a hexahedron structure, composed by
 six dimers with the C monomers pointing to the center of the struture. 
 These structures are shown in figures~\ref{fig5}(a) and ~\ref{fig5}(b)
 for the temperature $T=0.5$ and density
 $\rho=0.10$ (tetrahedral) and $\rho=0.15$ (hexahedron). 
 Increasing the density, as $\rho=0.20$, more dimers
 attach to the micelles, and the structure changes from a hexahedron to a 
 spherical micelle.  This structure was observed inside
 the region limited by the magenta line in figure~\ref{fig4}(a).
 Finally, at even higher densities, as $\rho=0.30$, the high pressure 
 changes the micelles shape from spherical to elongated micelles,
 as shown in figure~\ref{fig5}(c) and \ref{fig5}(d). The transition between 
 tetrahedral -- hexahedron -- spherical micelle can be understand
 by the fact that increasing the density (and consequently the pressure) 
 more C monomers will be attached in a micelle, changing
 the structure shape. However, the spherical to elongated micelle transition 
 is lead by the competition between the two length scales
 in the anomalous potential. The RDF between A monomers, displayed in figure~\ref{fig5}(e), 
 shows that, in the density range were the first three species of micelles were observed,
 all A monomers are at the second length scale, $r_{ij}\approx2.0$, or further, 
 while the first characteristic length scale, at
 $r_{ij}\approx1.2$, do not have any occupancy. However, when the system 
 is structured in elongated micelles the first length
 scale sharply increases, showing that the particles moved from the 
 second length scale to the first one.
 
  The presence of the self-assembled phases not present
in the A-B and A-A cases are not surprising. They 
are present in the  usual hydrophobic-hydrophilic janus
particles. These phases are a result of the competition
between the repulsion of the core-softened potential
and the attraction of the LJ. As in the 
competing interaction models, all the transitions
including the transition between the fluid to structured phases
are first-order.
 
\section{Conclusion}
\label{Conclu}

In this work we have analyzed the pressure versus temperature phase diagram of a Janus dumbbells
system comparying
the effects of the 
competition between two length scales potential 
and an repulsive and attractive potentials. Each nanoparticle was composed by one anomalous monomer modeled 
with a two length potential and one monomer modeled as a standard LJ particle.

We found that when the only competition is between
the two length scales, the system
only shows the lamellar phase defined by the core-softened potential.
In this case as 
in the pure CS dumbbell, density and diffusion anomalies are present in the pressure versus temperature phase diagram.
 In the case of he CS-attractive LJ
dumbbell, the attraction affects the competition that lead to
waterlike anomalies. As consequence, the density anomaly vanishes and 
the diffusion anomaly region shrinks. Also, the models shows a 
rich variety of micelles in self-assembly process similar 
to the behavior observed in the hydrophobic-hydrophilic janus dumbbell.
Due to the attraction between C monomers and the two length scales competition 
between A monomers the nanoparticles can assembly to tetrahedral, hexahedron, spherical
or elongated micelles.

Our results indicated that is possible to create distinct colloidal 
particles that will have waterlike anomalies and different micellar
conformation. Furthers investigations of Janus dumbbells, including
distinct LJ well depthness, monomers size and separation, are currently
in progress.

\section{Acknowledgments}

We thank the Brazilian agencies CNPq, INCT-FCx, and Capes for the financial support.
We also thank to TSSC - Grupo de Teoria e Simula\c{c}\~{a}o em Sistemas Complexos 
at UFPel for the computational time in Satolep cluster.
%
%

\end{document}